\documentclass[aps,prd,twocolumn,showpacs,draft,floatfix]{revtex4}
\usepackage{amsbsy}
\usepackage{amsmath}
\usepackage{epsf}
\usepackage{dcolumn}
\begin{document}

\title{Snyder noncommutative space-time from two-time physics}

\author{Juan M. Romero}
\email{sanpedro@nuclecu.unam.mx}
\affiliation{Instituto de Ciencias Nucleares, UNAM, Apartado Postal 70-543,
 M\'exico DF, M\'exico}
\author{Adolfo Zamora}
\email{zamora@nuclecu.unam.mx}
\affiliation{Instituto de Ciencias Nucleares, UNAM, Apartado Postal 70-543,
 M\'exico DF, M\'exico}

\date{\today}

\begin{abstract}
We show that the two-time physics model leads to a mechanical system 
with Dirac brackets consistent with the Snyder noncommutative space. 
An Euclidean version of this space is also obtained and it is shown
that both spaces have a dual system describing a particle in a
curved space.
\end{abstract}


\maketitle

\section{Introduction}

Inspired by a conformal field theory, R.~Marnelius \cite{Mr:gnus}
built a classical mechanics model having the conformal group as the
global symmetry and the symplectic group $S_{P}(2)$ as the local one.
This model has interesting unusual properties. One of them is
that it must have two time coordinates; that is why it is normally
called the two-time physics (2T) model. By imposing different gauge 
conditions on this, one can obtain systems such as the 
relativistic particle with mass and the massless free particle
in the $AdS$ space-time. In this sense the 2T model can be used as
a toy model for unification. Supersymmetric extensions of the 2T
model can be found in Ref. \cite{super:gnus}. Recently, I. Bars
and co-workers reinvented the 2T model in string theory 
\cite{Ibars:gnus} and carried out several extensions in
different contexts (see Refs. \cite{Bars:gnus,Ibars1:gnus}
and references therein). Ref. \cite{Otros:gnus} also deals
with the same problem. In another work, M.~Montesinos, C.~Rovelli 
and T.~Thiemann proposed a classical mechanics model simulating
the gauge structure of general relativity. In this, the 
gauge group is $SL(2,R)$ \cite{merced:gnus} and, since $SL(2,R)$
is isomorphic to $S_{P}(2)$, this model is analogous to the 2T.
As it is, the 2T model has several interesting
properties one would like to see in a realistic model.\\

From different results in string theory \cite{witten:gnus},
the possibility that the space-time at short distances is
noncommutative has been extensively studied recently.  R.~Snyder 
\cite{Snyder:gnus} investigated these ideas first and built a
noncommutative Lorentz invariant discrete space-time: the so 
called Snyder space. Contrarily to the noncommutative spaces
from string theory, in Snyder space the noncommutativity 
depends on the space-time. After the work of Kontsevich
\cite{kontsevich:gnus}, Snyder-like spaces in the sense of
noncommutativity have attracted great attention. Snyder
space is also interesting because it can be mapped to
the $k$-Minkowski space-time \cite{kowalski:gnus}.
This space-time is a realization of the ``Doubly Special 
Relativity'' theory, which is a new proposal to deal with 
quantum gravity phenomena \cite{amelino:gnus}. An important 
result from loop quantum gravity, in addition, is that it 
leads to discrete geometric quantities \cite{Livine:gnus}, 
and in this sense the discreteness of Snyder space becomes 
also attractive.

We show in this investigation how, by imposing an alternative 
gauge condition on the 2T model, one gets to a mechanical system 
with Dirac brackets consistent with the commutation rules of the 
Snyder noncommutative space. Using other gauge conditions, we also 
show that an Euclidean version of the Snyder space can be obtained. 
Then, by exploiting the symmetries of the Hamiltonian, we conclude 
that each system has a dual. For the Snyder space the dual system 
is the massless particle in the $AdS$ space, but for the Euclidean 
Snyder it is the non-linear sigma model in one dimension.\\

The work in this paper is organized as follows. In Section $2$ 
a brief introduction to the 2T model is provided. Then, in 
Section $3$ the gauge conditions to get to the Snyder space are 
given. The analogous conditions, but to obtain the Euclidean Snyder
space are determined in Section $4$. In Section $5$ we show that
both of these spaces have a dual system; and finally in Section 
$6$ we summarize our results.\\

\section{The 2T model}
Let us first review some properties of the 2T action and its 
symmetries.

\subsection{2T action}

The action for the 2T model is defined as the Hamiltonian action
\begin{equation}
S=\int_{\tau_{1}}^{\tau_{2}} d\tau \bigg [\dot X\cdot P-
\left(\lambda^{1}\frac{1}{2}P^{2}+\lambda^{2}X\cdot P+
\lambda^{3}\frac{1}{2}X^{2}\right)\bigg ]\,,\label{eq:2s}
\end{equation}
with the Hamiltonian given by
\begin{equation}
H_{2T}=\left(\lambda^{1}
\frac{1}{2}P^{2}+\lambda^{2}X\cdot P+\lambda^{3}
\frac{1}{2}X^{2}\right)\,, \label{eq:H2T}
\end{equation}
where $\lambda^{1},\lambda^{2},\lambda^{3}$ are Lagrange
multipliers. From this, one can obtain the equations of motion
\begin{eqnarray}
&\dot X^{M} &\!\!\!=\lambda^{1}P^{M}+\lambda^{2}X^{M}\,,\label{eq:igo1}\\
&\dot P^{M} &\!\!\!=-\lambda^{2}P^{M}-\lambda^{3}X^{M}\,,\label{eq:igo2}\\
&P^{2} &\!\!\!\approx X^{2}\approx X\cdot P\approx 0\,,
\label{eq:igo3}
\end{eqnarray}
where the symbol of weak equivalence $(\approx)$ has been used in
the sense of Dirac \cite{D1:gnus,Te:gnus}. Now, by defining
\begin{equation}
\phi_{1}=\frac{1}{2}P^{2}\,,\quad
\phi_{2}=X\cdot P\,,\quad\phi_{3}=\frac{1}{2}X^{2}\,, \label{eq:2gen}
\end{equation}
and considering that the Poisson brackets are given by: 
$\{X_{M},P_{N}\}=\eta_{MN}$, and zero otherwise, with $\eta_{MN}$ 
being a flat metric, it can be seen that the algebra
\begin{equation}
\{\phi_{2},\phi_{3}\}=-2\phi_{3}\,,\quad
\{\phi_{2},\phi_{1}\}=2\phi_{1}\,,\quad
\{\phi_{1},\phi_{3}\}=-\phi_{2}\,, \label{eq:Merced}
\end{equation}
holds. That is, all three constraints are first class.
Eq. (\ref{eq:Merced}) represents the Lie algebra of
the $S_{P}(2)$ group which is formed by the $2\times 2$
matrices with determinant one. If one redefines variables
as
\begin{equation}
H_{1}=\phi_{1}\,, \quad H_{2}=-\phi_{3}\,, \quad D=\phi_{2}\,,
\end{equation}
the Lie algebra of the $SL(R,2)$ is obtained. This has been
already proposed as a toy model simulating the gauge group of 
general relativity \cite{merced:gnus}.\\

Now, if we consider the Euclidean or Minkowski metrics as the 
background space, the surface defined by Eq. (\ref{eq:igo3}) 
is trivial. Therefore, the simplest metric giving a non-trivial
surface is the flat metric with two time coordinates. Throughout 
this work we will assume this metric only.
If the configuration space has dimensionality $D=d+2$, a
flat metric $\eta_{MN}$ with signature
\begin{equation}
{\rm sig}(\eta )=(-,-,+,\cdots,+)\,, \label{eq:signa}
\end{equation}
must be used. The coordinates of the phase space can be taken as
\begin{eqnarray}
&X^{M}&\!\!\!=(X^{0\prime},X^{1\prime}, X^{0}, X^{i})\,, \nonumber\\
&P^{M}&\!\!\!=(P^{0\prime},P^{1\prime}, P^{0},P^{i})\,,\quad (i=1,\dots,d-1)\,,
\label{eq:coor}
\end{eqnarray}
where the zeroes are associated with the time coordinates.\\

In principle the phase space of the system has $2(d+2)$
independent coordinates. However, as there are three first-class
constraints, six degrees of freedom must be subtracted. Therefore,
there are $2(d-1)$ effective degrees of freedom and so the configuration
space has $(d-1)$ independent coordinates.

\subsection{Symmetries}

The equations of motion (\ref{eq:igo1}) and (\ref{eq:igo2}) can
be rewritten as
\begin{equation}
\frac{d}{dt}\left(
\begin{array}{c}
X^{M} \\
P^{M}  \\
\end{array} \right)=
A(t)
\left(
\begin{array}{c}
X^{M} \\
P^{M}  \\
\end{array} \right)\,,
\label{eq:pao}
\end{equation}
with
$
A(t)=\left(
\begin{array}{cc}
\lambda^{2} & \lambda^{1}\\
-\lambda^{3} & -\lambda^{2}
\end{array} \right)\,.
$
By performing a gauge transformation with an arbitrary
matrix of $Sp(2)$,
\begin{equation}
\left(
\begin{array}{c}
\bar X^{M} \\
\bar  P^{M}  \\
\end{array} \right)=
U(t)
\left(
\begin{array}{c}
X^{M} \\
P^{M}  \\
\end{array} \right)\,,\label{eq:ramona}
\end{equation}
where
$
U(t)=\left(
\begin{array}{cc}
a & b\\
c & d
\end{array} \right)
,\quad ad-bc=1\,,
$
one gets to the transformed equations of motion
\begin{equation}
\frac{d}{dt}\left(
\begin{array}{c}
\bar X^{M} \\
 \bar P^{M}  \\
\end{array} \right)=
\bar A(t)
\left(
\begin{array}{c}
\bar X^{M} \\
\bar P^{M}  \\
\end{array} \right)\,,
\end{equation}
where 
\begin{equation}
\bar A(t)=U(t)A(t)U(t)^{-1}-U(t)\frac{dU(t)^{-1}}{dt}\,. \label{eq:abel}
\end{equation}
It can be easily seen that $A(t)$ transforms as a connection under
the gauge transformation $U(t)$ and that the equations of motion
(\ref{eq:pao}) are invariant under this gauge transformation
as well.\\

Now, the action in Eq. (\ref{eq:2s}), when rewritten in terms
of the transformed variables, takes the form
\begin{eqnarray}
S &\!\!\!=\!\!\!&\int_{\tau_{1}}^{\tau_{2}} d\tau 
\bigg [\dot X\cdot P-\left(\lambda^{1}
\frac{1}{2}P^{2}+\lambda^{2}X\cdot P+\lambda^{3}
\frac{1}{2}X^{2}\right)\bigg ]\nonumber\\
&\!\!\!=\!\!\!&\int_{\tau_{1}}^{\tau_{2}} d\tau \bigg 
[\frac{d \bar X}{d\tau}\cdot \bar P-\left(\bar\lambda^{1}
\frac{1}{2}\bar P^{2}+\bar\lambda^{2}\bar X\cdot \bar P+\bar\lambda^{3}
\frac{1}{2}\bar X^{2}\right)\nonumber\\
& &{} + \frac{d}{d\tau}\left(-ab \frac{1}{2}\bar P^{2}+bc \bar X\cdot 
\bar P-dc\frac{1}{2}\bar X^{2}\right) 
\bigg ]\,, \label{eq:elisa}
\end{eqnarray}
where $\bar \lambda^{i}$ is given by Eq. (\ref{eq:abel}). Thus, up to a
boundary term, the action in Eq. (\ref{eq:2s}) is invariant under the
gauge transformations (\ref{eq:ramona}) and (\ref{eq:abel}).
On the other hand, the quantities $X\cdot P$, $ X^{2}$, $ P^{2}$ and 
$\dot X\cdot P$ are clearly invariant under global transformations
$\Lambda$ that satisfy
\begin{equation}
\Lambda^{T}\eta\Lambda =\eta\,,
\end{equation}
with the signature $\eta$ defined in Eq. (\ref{eq:signa}). Thus, the
action in Eq. (\ref{eq:2s}) is invariant under global transformations
of $SO(2,d)$. It can be shown that in phase space the generators of
this symmetry are
\begin{equation}
L^{MN}=X^{M} P^{N}-X^{N} P^{M}\,,\label{eq:gero}
\end{equation}
which satisfy the conformal algebra \cite{Maldacena:gnus} and are
conserved quantities. Moreover, they satisfy $\{L^{MN},\phi_{i}\}=0$,
i.e. they are gauge invariant.\\

\section{Snyder space}

Let us now consider the gauge conditions to get the Snyder space
\begin{equation}
P_{1^{\prime}}=L={\rm const.}\,, \qquad X_{1^{\prime}}=0\,.\label{eq:s1}
\end{equation}
Substituting them into the equations of motion (\ref{eq:igo1}) and
(\ref{eq:igo2}) we obtain
\begin{equation}
\lambda^{2}=\lambda^{1}=0\,.
\end{equation} 
By using Eq. (\ref{eq:igo3}) it can be seen that the independent
reduced equations of motion are
\begin{eqnarray}
&\dot X^{\mu}&\!\!\!=0\,,\\
&\dot P_{\mu}&\!\!\!=-\lambda^{3}X_{\mu}\,,\\
&\phi_{3}&\!\!\!=\frac{1}{2}G_{\mu\nu}X^{\mu}X^{\nu}\approx 0,\quad 
G_{\mu\nu}= \left(\eta_{\mu\nu}-
\frac{P_{\mu}P_{\nu}}{P_{\alpha}P^{\alpha}+L^{2}}\right),
\end{eqnarray} 
with the dependent variables being
\begin{equation}
P_{0^{\prime}}=\sqrt{P_{\mu}P^{\mu}+L^{2}}\,,\quad  
X_{0^{\prime}}=\frac{P_{\mu}X^{\mu}}{\sqrt{P_{\mu}P^{\mu}+L^{2}}}\,.
\end{equation}
After performing an integration by parts and substituting the
dependent variables into Eq. (\ref{eq:2s}) one obtains 
\begin{equation}
S=\int d\tau \left[-G_{\mu\nu}X^{\mu}\dot P^{\nu}-\frac{\lambda^{3}}{2}
G_{\mu\nu}X^{\mu}X^{\nu}\right]\,. 
\end{equation} 
To quantize this system with the canonical formalism, the Dirac 
brackets \cite{D1:gnus,Te:gnus} must be constructed. In this process 
the Dirac brackets are replaced by commutators. Now, let us consider
\begin{eqnarray}
&\chi_{1}&\!\!\!=P_{1^{\prime}}-L\,,\\
&\chi_{2}&\!\!\!=X_{1^{\prime}}\,,\\
&\chi_{3}&\!\!\!=P\cdot X\,,\\
&\chi_{4}&\!\!\!=\frac{1}{2}P^{2}\,,\\
&\phi&\!\!\!=\frac{1}{2}X^{2}\,. \label{eq:phi}
\end{eqnarray}
A straightforward calculation shows that Eq. (\ref{eq:phi}) is a 
first-class constraint while the others are second class. For the 
later ones we find
\begin{equation}
C_{\alpha\beta}\approx\{\chi_{\alpha},\chi_{\beta}\} \approx \left(
\begin{array}{rrrr}
 0       & -1     & -L  & \quad 0  \\
 1       &  0     &  0  &       L  \\
 L       &  0     &  0  &       0   \\
 0       & -L     &  0  &       0         
\end{array}
\right)\,, \label{eq:Definida}
\end{equation}
from which,
\begin{equation}
C^{\alpha\beta} \approx -\frac{1}{L}\left(
\begin{array}{rrrr}
 0      &  0      & -1           & 0  \\
 0      &  0      &  0           & 1 \\
 1      &  0      &  0           & - \frac{1}{L}  \\
 0      & -1      &  \frac{1}{L} & 0         
\end{array}
\right)\,.
\end{equation}
In general, given two functions $A$ and $B$ in phase space,
the Dirac brackets are defined as
\begin{equation}
\{A, B\}^{*}=\{A,B\}-\{A,\chi_{\alpha}\}C^{\alpha\beta}\{\chi_{\beta},B\}\,.
\end{equation}
In particular, for the phase space coordinates
\begin{eqnarray}
&\{X_{\mu}\,,X_{\nu}\}^{*}&\!\!\!=\frac{1}{L^{2}}\left(X_{\mu}P_{\nu}-
X_{\nu}P_{\mu}\right)\,,\\
&\{X_{\mu}\,,P_{\nu}\}^{*}&\!\!\!=\eta_{\mu\nu}+\frac{1}{L^{2}}P_{\mu}P_{\nu}\,,\\
&\{P_{\mu}\,,P_{\nu}\}^{*}&\!\!\!=0\,.
\end{eqnarray}
These Dirac brackets are the classic version of the commutation rules
of Snyder space \cite{Snyder:gnus}. Therefore, after quantizing the
system we have the Snyder space as the background.

Now, by defining ${\bf X_{\mu}}=G_{\mu\nu}X^{\nu}$, the pair
$({\bf X_{\mu}}, P_{\mu})$ satisfies the Dirac brackets
\begin{equation}
\{{\bf X}_{\mu},{\bf X}_{\nu}\}^{*}=0\,,\quad 
\{{\bf X}_{\mu},P_{\nu}\}^{*}=\eta_{\mu\nu}\,,\quad
\{P_{\mu},P_{\nu}\}^{*}=0\,.
\end{equation}
That is, with the variables $({\bf X_{\mu}}, P_{\mu})$,
the usual Poisson brackets are obtained. Nevertheless,
at the quantum level the definition of ${\bf X_{\mu}}$
is ambiguous.

\section{Euclidean Snyder space}

Other gauge conditions from which a noncommutative space can
be obtained are
\begin{equation}
\chi_{1}=P_{0}-1=0\,, \qquad \chi_{2}=X_{0}=0\,. \label{eq:coco}
\end{equation}
For these the independent equations of motion are
\begin{eqnarray}
&\dot X^{i}&\!\!\!=0\,,\\
&\dot P^{i}&\!\!\!=-\lambda^{3} X^{i}\,,\\
&\phi_{3}&\!\!\!=g_{ij}X^{i}X^{j}\approx 0\,, \quad 
g_{ij}=\left(\delta_{ij}+\frac{P_{i}P_{j}}{1-P_{k}P^{k}}\right)\,, 
\end{eqnarray}
and, as can be easily seen, the second-class constraints are 
given by
\begin{eqnarray}
&\chi_{1}&\!\!\!=P_{0}-1\,,\\
&\chi_{2}&\!\!\!=X_{0}\,,\\
&\chi_{3}&\!\!\!=P\cdot X\,,\\
&\chi_{4}&\!\!\!=\frac{1}{2}P^{2}\,.
\end{eqnarray}
From a straightforward calculation it can be observed that in this
case the matrix $C_{\alpha\beta}\approx\{\chi_{\alpha},\chi_{\beta}\}$
is minus the matrix in Eq. (\ref{eq:Definida}) with $L=1$. Using this,
we find for the phase space coordinates
\begin{eqnarray}
&\{X_{i}\,,X_{j}\}^{*}&\!\!\!=-\left (X_{i}P_{j}-X_{j}P_{i}\right)\,,\\
&\{X_{i}\,,P_{j}\}^{*}&\!\!\!=\delta_{ij}-P_{i}P_{j}\,,\\ 
&\{P_{i}\,,P_{j}\}^{*}&\!\!\!=0\,.
\end{eqnarray}
Thus, after quantizing the reduced system, a noncommutative
space in the coordinates is obtained.\\

By defining the variable ${\bf X}_{i}=g_{ij}X^{j}$, it can be
seen that the pair $({\bf X}_{i}, P_{j})$ satisfies 
\begin{equation}
\{{\bf X}_{i},{\bf X}_{j}\}^{*}=0\,,\quad
\{{\bf X}_{i},P_{j}\}^{*}=\delta_{ij}\,,\quad  
\{P_{i},P_{j}\}^{*}=0\,.
\end{equation}
Now, as the only gauge transformations permitted are of the type
of Eq. (\ref{eq:ramona}), there is no gauge transformation which
takes the Snyder space to this system. In this sense they are
different physical systems.

\section{$O(d+1)$ non-linear sigma model in one dimension}

It can be seen that the Hamiltonian $H_{2T}$ from Eq. (\ref{eq:H2T})
is invariant under the transformations
\begin{equation}
(X^{M}, P^{M})\to (P^{M}, X^{M})\,,\quad 
(\lambda^{1},\lambda^{2},\lambda^{3})\to 
(\lambda^{3},\lambda^{2},\lambda^{1})\,. 
\end{equation}
This symmetry implies that if we impose gauge conditions 
and then the $X$s and $P$s are swapped, we obtain analogous 
reduced systems. Notice, however, that the physical interpretation 
of each one is different. As an example, by performing this swap 
in the gauge conditions of the Euclidean Snyder space from Eq. 
(\ref{eq:coco}), one obtains
\begin{equation}
\chi_{1}=X_{0}-1=0 \qquad{\rm and}\qquad \chi_{2}=P_{0}=0\,.
\label{eq:nopasa}
\end{equation}
In this case the independent reduced equations of motion are
\begin{eqnarray}
&\dot X^{i}&\!\!\!=\lambda^{1}P^{i}\,, \qquad (i=1,\dots,d)\label{eq:perez}\\
&\dot P^{i}&\!\!\!=0\,,\label{eq:monarres}\\
&\phi_{1}&\!\!\!=\tilde g_{ij}P^{i}P^{j}\approx 0, \quad \tilde g_{ij}=\left(\delta_{ij} +
\frac{X_i X_j}{1-X_k X^k}\right), \label{eq:monarres1}
\end{eqnarray}
with dependent variables given by
\begin{equation}
X^{0'}=\sqrt{X^{i}X_{i}-1}\,,\quad 
P^{0'}=\frac{(P^{i}X_{i})}{\sqrt{X^{i}X_{i}-1}}\,.
\label{eq:america2}
\end{equation}
Now, by rewriting Eq. (\ref{eq:2s}) in terms of the independent variables
we obtain
\begin{equation}
S= \int _{\tau_{1}} ^{\tau_{2}} d\tau \left[\tilde g_{ij}\dot X^{i}
P^{j}-\frac{\lambda^{1}}{2}\tilde g_{ij} P^{i}P^{j}\right]\, .
\label{eq:tardesy}
\end{equation}
From this expression one 
gets to the equations of motion (\ref{eq:perez})--(\ref{eq:monarres1}).
Now, substituting Eq. (\ref{eq:perez})
into Eq. (\ref{eq:tardesy}) and eliminating $P_{i}$ as a
dynamic variable we get
\begin{equation}
S=\frac{1}{2}\int _{\tau_{1}} ^{\tau_{2}} d\tau \frac{\tilde g_{ij}\dot 
X^{i}\dot X^{j}}{\lambda^{1}}\,.\label{eq:mesias}
\end{equation}
Eq. (\ref{eq:mesias}) could be interpreted as the action of a massless
free particle in a space with metric $\tilde g_{ij}$, but non-relativistic 
massless particles are not natural. In a better interpretation, for 
$\lambda^{1}=1$, this equation represents the action of the
$O(d+1)$ non-linear sigma model in one dimension \cite{Zin:gnus}.\\

The Dirac brackets for the phase space coordinates, in this case, are
\begin{eqnarray}
&\{X_{i}\,,X_{j}\}^{*}&\!\!\!=0\,, \label{eq:LL}\\
&\{X_{i}\,,P_{j}\}^{*}&\!\!\!=\delta_{ij}-X_{i}X_{j}\,,\\ 
&\{P_{i}\,,P_{j}\}^{*}&\!\!\!=- \left(X_{i}P_{j}-X_{j}P_{i}\right)\,
 \label{eq:LLL}.
\end{eqnarray}
However, using the coordinates $(X_{i}, \bar{\bf P}_{i}=\tilde g_{ij}P^{j})$
one gets to
\begin{equation}
\{X_{i}\,,X_{j}\}^{*}=0\,, \quad 
\{X^{i}\,,{\bf \bar P}_{j}\}^{*}=\delta^{i}_{j}\,,\quad  
\{{\bf \bar P}_{j}\,,{\bf \bar P}_{j}\}^{*}=0\, \label{eq:rech}.
\end{equation}
In terms of the variables $(X_{i}, \bar{\bf P}_{i})$, the action
of Eq. (\ref{eq:tardesy}) becomes
\begin{equation}
S= \int _{\tau_{1}} ^{\tau_{2}} d\tau \left[\dot X^{i}
{\bf P}_{i}-\frac{\lambda^{1}}{2}\tilde g^{ij} {\bf P}_{i}{\bf P}_{j}
\right]\, .
\end{equation}
Thus, this system can be thought of: either a particle in an
Euclidean metric with a deformed Poisson structure, Eqs.
(\ref{eq:LL})--(\ref{eq:LLL}), or as a particle in the
metric $\tilde g_{ij}$ with the standard Poisson structure,
Eq. (\ref{eq:rech}). A similar interpretation can be given 
to the systems presented in sections $3$ and $4$.\\

By performing the change of variables $(X,P)\to (P,X)$ in the gauge
conditions for the Snyder space, Eq. (\ref{eq:s1}), one gets to the
gauge conditions
\begin{equation}
X_{1^{\prime}}=L={\rm const.}\,, \qquad P_{1^{\prime}}=0\,.
\label{eq:s11}
\end{equation}
In Refs. \cite{Mr:gnus} and \cite{Ibars1:gnus} it is shown that,
using the conditions from Eq. (\ref{eq:s11}), the massless
particle in the $AdS$ space is obtained. This can also be easily 
verified by repeating the calculation using the conditions
of Eq. (\ref{eq:nopasa}). \\

It is remarkable that in the 2T model both dynamics in noncommutative
spaces have as dual a dynamics in a curved space-time.

\section{Summary}

In this work we study a mechanical system with two times and
gauge freedom called the two-time physics. It is shown that
considering a particular gauge one gets a mechanical system
with Dirac brackets consistent with the commutation rules
of the Snyder noncommutative space. Using other gauge conditions
an Euclidean version of the Snyder space is obtained. By 
exploiting a symmetry of the Hamiltonian we show that these
noncommutative systems have a dual system. For the Snyder
space, the dual is a massless particle in the $AdS$ space,
while for the Euclidean Snyder the dual is the non-linear
sigma model in one dimension.

\begin{acknowledgments}
The authors would like to thank J.~D.~Vergara for discussions.
\end{acknowledgments}

\end{document}